\documentstyle[aps,eqsecnum,epsf,subfigure]{revtex}

\begin{document} 
\bibliographystyle{prsty} 
\newcommand {\gtsim } {\,\vcenter{\hbox{$\buildrel\textstyle>\over\sim$}}\,}
\newcommand {\BE} {\begin{equation}}
\newcommand {\EE} {\end{equation}}
\newcommand  {\sign }  {\,{\rm sign} \,}

\draft

\title{Wound-up phase turbulence in 
the Complex Ginzburg-Landau Equation} 

\author{ R. Montagne\footnote{on leave from {U}niversidad de la {R}ep{\'u}blica
({U}ruguay).}$^{\ddag\S}$, 
E. Hern\'andez-Garc\'\i a$^{\ddag\S}$, A. Amengual$^\ddag$, and M. San
Miguel$^{\ddag\S}$ 
} 
\address{$^\ddag$Departament de F\'\i sica, Universitat
de les Illes Balears, E-07071 Palma de Mallorca, Spain \\
$^\S$ Instituto Mediterr\'aneo de Estudios Avanzados, IMEDEA\footnote{URL:
http://www.imedea.uib.es/Nonlinear} (CSIC-UIB),
 E-07071 Palma de Mallorca, Spain 
} 

\date{December 16, 1996} 

\maketitle
\begin{abstract}

We consider phase turbulent regimes with nonzero winding number
in the one-dimensional Complex Ginzburg-Landau equation. We find that 
phase turbulent states
with winding number larger than a critical one are only transients and decay to
states within a range of allowed winding numbers. The analogy with the Eckhaus 
instability for non-turbulent waves is stressed. The transition
from phase to defect turbulence is interpreted as an ergodicity breaking 
transition which occurs when the 
range of allowed winding numbers vanishes. We explain the states reached at long
times in terms of three basic states, namely {\sl quasiperiodic} states, {\sl 
frozen turbulence} states, and {\sl riding turbulence} states. Justification and
some insight into
them is obtained from an analysis of a phase equation for nonzero winding
number: rigidly moving solutions of this equation, which correspond to 
quasiperiodic and frozen turbulence states, are understood in terms
of periodic and chaotic solutions of an associated system of ordinary
differential equations. A short 
report of some of our results has been published in [{\sl Montagne et al., 
Phys. Rev. Lett. {\bf 77}, 267 (1996)}]. 
\end{abstract} 

\vskip 0.4cm

\section{Introduction}
\label{intro}

\subsection{The complex Ginzburg-Landau equation and its phase diagram}

Spatio-temporal complex dynamics
\cite{CrossHohenberg,CrossHohenberg2,dennin96} is one of the present focus of
research in nonlinear phenomena.  This subject lies at the intersection of
two important lines of thought: on the one hand the generalization of the 
ideas of dynamical systems theory to high dimensional
situations\cite{egolf194,bohr194,bohrbook},  and on the other the 
application of some concepts and tools developed in the field of
statistical  mechanics, specially in the study of phase transitions, to the
analysis of  complex nonequilibrium systems 
\cite{chate1,hohenberg89,ciliberto2}.   

An important effort has been devoted to the
characterization of different dynamical states and transitions among
them for model equations such as the Complex Ginzburg-Landau Equation 
(CGLE)
\cite{CrossHohenberg,egolf194,chate1,montagne96b,chate2,chate3,janiaud1,hohenbergsaarloos,chate8,egolf195,montagne96a}.
  The CGLE is an equation for a complex field $A({\bf x},t)$:  
\begin{equation} 
 \partial_{t} A = A + (1 + {\it i} c_1 ) \nabla^{2} A - 
 (1 + {\it i} c_2 ) \mid A \mid^{2} A \ . 
 \label{cgle} 
\end{equation} 
$A({\bf x},t)$ represents the slowly varying, in space and time,  complex
amplitude of the Fourier mode of zero  wavenumber when it has become unstable
through a Hopf bifurcation (the signs  used in (\ref{cgle}) assume it to be
supercritical). The CGLE is obtained universally when analyzing the dynamics
sufficiently close to the bifurcation point. In one dimensional
geometries, (\ref{cgle}) or a coupled set of similar equations  with 
additional group
velocity terms describe also the evolution of the amplitudes of
Hopf-bifurcated traveling waves
\cite{CrossHohenberg,hohenbergsaarloos,toni96}. Binary fluid convection
\cite{kolodner95}, transversally extended lasers \cite{coullet,maxi95},
chemical turbulence\cite{kuramoto74,kuramoto81}, bluff body
wakes \cite{provansal94}, the motion of bars in the bed of rivers 
\cite{schielen93},
and many other systems have been described by the CGLE in the appropriate
parameter range. We will restrict ourselves in this paper to the
one-dimensional case, that is $A=A(x,t)$, with $x \in [0,L] $.  As usual, we
will use periodic boundary conditions in $x$.

\begin{figure}[H] 
\begin{center}
\epsfysize=60mm
\vspace{60mm}
{\epsffile{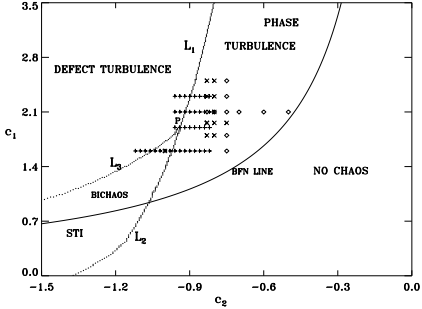}}
\vspace{-60mm}
\end{center}
\caption{\label{fig1} Regions of the parameter space $[c_{1}, c_{2}]$
for the CGLE displaying different kinds of regular and chaotic behavior.
Lines $L_1, L_3$ were determined
in\protect\cite{chate1,chate2,chate3}. See the text in Section II for the 
explanation of the different symbols. } 
\end{figure}

The one-dimensional Eq. (\ref{cgle}) has traveling wave (TW) solutions 
\BE 
\label{planew}
A_k=\sqrt{1-k^2}e^{i(kx-\omega_k t)}, \ \omega_k = c_2 + (c_1 -c_2) k^2 
\EE
with $k \in [-1,1]$. When $1+c_1c_2>0$ there is a range of wavenumbers
$[-k_E, k_E]$ such that TW solutions with wavenumber in this
range are linearly stable. Waves with $k$ outside this range display a
sideband instability (the Eckhaus instability 
\cite{CrossHohenberg,janiaud1,montagne96e}). 
The limit of this range, $k_E$, vanishes as the quantity $1+c_1c_2$
approaches zero, so that the range of stable traveling waves vanishes by
decreasing $1+c_1c_2$. The line $1+c_1c_2=0$, is the 
Benjamin--Feir--Newell line\cite{bf1,bf2}, labeled BFN in Fig.\ref{fig1}.
Above that  line, where $1+c_1c_2<0$, no traveling wave is stable and different
turbulent  states exist. A major step towards the analysis of phases and
phase transitions in (\ref{cgle}) was the numerical construction in 
\cite{chate1,chate2,chate3}   of a phase diagram that shows which type of
regular or chaotic behavior occurs in different regions of the parameter
space $[c_1,c_2]$. Fig. \ref{fig1} has  been constructed from the data in
\cite{chate1,chate2,chate3}. Above the BFN line, three types of
turbulent behavior are found, namely {\sl phase turbulence} (PT), {\sl
defect} or {\sl amplitude turbulence} (DT), and  {\sl bichaos} (BC).  
~

Phase turbulence is a state in which $A(x,t)=|A|e^{i\varphi}$ evolves
irregularly but  with its modulus always far from $|A|=0$. Since the
modulus never vanishes,  periodic boundary conditions enforce the {\sl
winding number} defined as  
\begin{equation}
\label{winding}
\nu \equiv \frac{1}{2 \pi} \int_0^L \partial_x \varphi dx
\end{equation}
to be a constant of motion, fixed by the initial condition. $\nu$ is 
always an integer because of 
periodic boundary conditions . The quantity
$\bar k \equiv 2 \pi \nu /L$ can be thought of as an {\sl average} or 
{\sl global 
wavenumber}.  To the left of line $L_1$ (region DT), in contrast, the 
modulus of
$A$ becomes  zero at some instants and places (called {\sl defects} or 
{\sl phase slips}). In such places the phase $\varphi$ becomes  undefined, so
allowing  $\nu$ to change its value during evolution. BC is a region in which
either PT,  DT, or spatial coexistence of both can be observed depending on
initial conditions.  It should be noted that chaotic states exist also below
the BFN line: To the left of the line $L_2$, a chaotic attractor called {\sl
Spatio Temporal Intermittency} (STI) coexists with the stable traveling waves 
\cite{chate2}. A diagram qualitatively similar to Fig. \ref{fig1} has also been
found for the two-dimensional CGLE \cite{chate6,chate9}. 
Despite the relevance of $\nu$ in the dynamics of the CGLE, most studies  of
the PT regime have only considered in detail the case of $\nu=0$. In fact the
phase diagram in Fig. \ref{fig1} was constructed
\cite{chate1,chate2,chate3}  using initial conditions that enforce
$\nu=0$. Apart from some limited
observations\cite{chate3,janiaud1,chate9},  systematic consideration of
the $\nu \neq 0$ ({\sl wound}) disordered phases  has started only 
recently \cite{montagne96b,torcini196,torcini296}. States with $\nu \neq 0$ 
are precisely the subject of the  present paper. 
\subsection{The PT-DT transition}

Among the regimes described above, the transition between PT and DT has
received special attention 
\cite{chate1,montagne96b,egolf195,torcini196,torcini296,sakaguchi2}. 
The PT regime is robustly observed for
the large but finite sizes and for the long but finite observation times
allowed by computer simulation, with the
transition to DT appearing at a quite well defined line ($L_1$ in Fig. 
\ref{fig1})
\cite{chate8,chate9}, but it is unknown if the PT state would persist 
in the thermodynamic limit $L\rightarrow\infty$. One possible scenario is that 
in a system large enough, and after waiting enough time, a defect would
appear somewhere, making thus the conservation of $\nu$ only an
approximate rule. In this scenario, a PT state is a long lived metastable 
state.  
In the alternative scenario, the one in which PT and the transition to DT 
persist
even in the thermodynamic limit, this transition would be a kind of ergodicity  
breaking transition \cite{montagne96b,palmer89} in which the system restricts 
its dynamics
to the small portion of configuration space characterized by a particular 
$\nu$.  DT would correspond to a ``disordered" 
phase and different ``ordered" phases in  the PT region would be classified 
by its value of $\nu$.  The idea of using a quantity 
related to $\nu$ as an order parameter \cite{montagne96b} has also been 
independently proposed in \cite{torcini196}. 

The question of which of the scenarios above is the appropriate one is not yet 
settled. Recent investigations seem to slightly favor the first
possibility   \cite{chate3,chate8,egolf195,chate9}. The most powerful
method in equilibrium statistical mechanics to distinguish true phase
transitions from sharp crossovers is the careful analysis of finite-size
effects \cite{barber83}. Such type of analysis has been carried out in
\cite{chate8,chate9} , giving some evidence (although not definitive)
that the PT state will not properly exist in an infinite system or,
equivalently,  that the $L_1$ line in Fig. \ref{fig1} approaches the BFN line as
$L\rightarrow\infty$.  Here we present another finite-size scaling
analysis, preliminarily commented in \cite{montagne96b}, based on the
quantity $\nu$ as an order parameter. Our result is
inconclusive, perhaps slightly favoring the vanishing of PT at large
system sizes. In any case, the PT regime is clearly observed in the largest
systems  considered and its characterization is of relevance for
experimental systems, that are always finite.  In this paper we characterize
this PT regime in a finite system as we now outline. 

\subsection{Outline of the paper}

We show that in the PT regime there is an
instability such that a conservation law for the winding number  occurs only
for $\nu$ within a finite range that depends on the point in parameter space.
PT states with too large $|\nu|$ are only  transients and decay to states
within a band of allowed winding numbers. Our results, presented in  Section
\ref{win},   allow a characterization of the transition from PT to DT in terms
of the range of conserved $\nu$: as one moves in parameter space, within the
PT regime and towards the DT regime, this range becomes smaller. The
transition is identified with the line in parameter space at which such
stable range vanishes. Analogies with known aspects of the Eckhaus and the
Benjamin-Feir instabilities are stressed. States with $\nu \neq 0$ found
in  the PT region of parameters at late times are of several types, and Section
\ref{asymstat} describes them in  terms of three \cite{montagne96b}
elementary {\sl wound} states. Section \ref{kawa} gives some insight into the
states numerically obtained by explaining them in terms of solutions of a phase
equation. In addition, theoretical predictions are made for such states. The
paper is closed with some final remarks. An Appendix explains our
numerical method.

\section{The winding number instability}
\label{win}

The dynamics of states with non-zero winding number and periodic boundary 
conditions has been studied numerically in the PT region of parameters. In
order to do so we have performed numerical integrations of Eq. (\ref{cgle}) 
in a number of points, shown in Fig. \ref{fig1}. Points marked as  $\Diamond$
correspond to parameter values where intensive statistics has been performed. 
The
points overmarked with $\times$ correspond to places where finite-size
scaling was analyzed. Finally the symbol  $+$ correspond to runs made in
order to determine accurately the PT-DT transition line ($L_1$). 
Our pseudospectral integration method is described in the Appendix. 
Unless otherwise stated, system size is $L=512$ and the spatial resolution 
is typically 512  modes, with some runs performed with up to 4096 modes to
confirm the results.  The initial condition is a traveling wave, with a desired
initial winding number $\nu_i$, slightly perturbed by a random noise of 
amplitude $\epsilon$. By this amplitude we specifically mean that a set of 
uncorrelated Gaussian numbers of zero mean and
variance  $\epsilon^2$ was generated, one number for each collocation point 
in the numerical lattice.  Only
results for $\nu_i > 0 $ are shown here. The behavior for $\nu_i < 0$ is
completely symmetrical. 

The initial evolution  is well described by the
linear stability analysis around the traveling wave 
\cite{janiaud1,hohenbergsaarloos,legath,montagne96e}. 
Typically, as seen from the evolution of the power spectrum, unstable 
sidebands initially grow. This growth stops when an intense competition 
among  modes close 
to the initial wave and to the broad sidebands is established. Configurations
during this early nonlinear regime are similar to the ones that would be called
{\sl riding turbulence} and described in Section \ref{asymstat}. At long times 
the system approaches one of several
possible dynamical states. In general, they can be understood in terms
of three of them, which are called basic states. In the next section
these final states are discussed.  When the initial winding number is above 
a critical value $\nu_c$, which depends on $c_1$ and $c_2$, there is a 
transient period between the early competition and the 
final state during which the winding number changes.

\begin{figure} 
\begin{center}
\vspace{-0mm}
\epsfysize=100mm
\epsffile{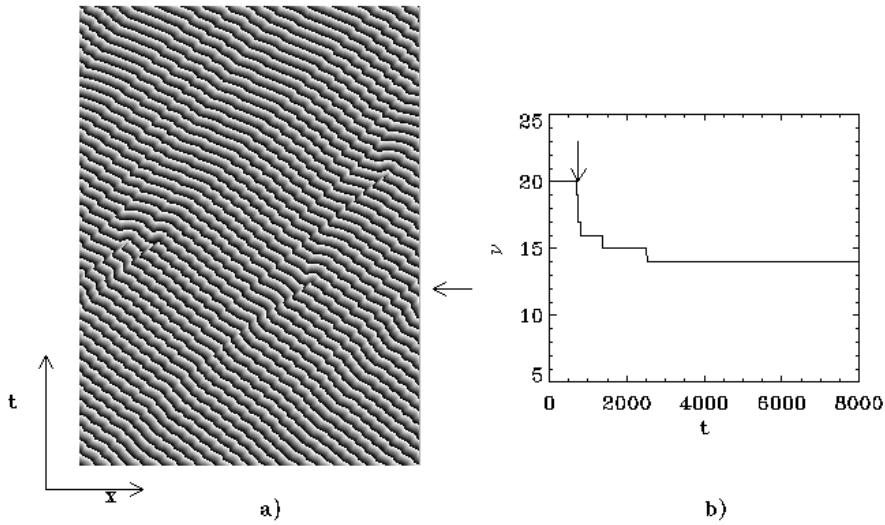}
\end{center}
\vspace{-0mm}
\caption{\label{fig2} a) Spatiotemporal evolution of the phase $\varphi(x,t)$ 
coded in grey levels with time running
upwards and $x$ in the horizontal direction.  The lighter
grey correspond  $\varphi(x,t) = - \pi$ and darker to 
$\varphi(x,t) =  \pi$. The time interval shown in the picture goes from 
$t = 500$  
to $1000$ time units of a total run of $10^4$. $c_1 =2.1$,  $c_2 = -0.60$,
 and the  initial condition was a TW with  $\nu_i = 20$ 
that decayed to $\nu_f = 14$. The arrow indicates the time at which $\nu$
begins to change.  b) The complete time evolution of the winding number for 
this 
 initial condition. }
\end{figure}

In Fig. \ref{fig2}a we show in grey levels the phase $\varphi(x,t)$ for a given
run with parameters $c_1=2.1$ and $c_2=-0.6$. The
space-time defects appear as dislocations in this representation. 
In Fig. \ref{fig2}b the
winding number has been plotted as a function of time. The winding number
changes  from the initial value $\nu_i=20$ to the final value $\nu_f=14$.  
The discrete jumps in $\nu$ are due to the integer nature of this quantity, and
they are smeared out when averages over several realizations  are
performed. The resemblance with the dynamics of the Eckhaus instability 
of regular waves is striking. In fact, since the changes in $\nu$ occur on top
of a chaotic wave, the analogy is stronger with the Eckhaus instability in
the presence of stochastic fluctuations \cite{emilio93,emilio92}.  In
the latter case a local wavenumber independent of position cannot be
defined because of noise, while for phase turbulent waves the disorder is
generated by the system dynamics.  Nevertheless in both
cases the configurations can be characterized by a global wavenumber 
such as $\bar k$ or $\nu$. 
The analogy is also instructive since it can be shown
\cite{emilio92,emilio91} that for the one--dimensional relaxational
dynamics considered in \cite{emilio93,emilio92,emilio91} 
(which is related to 
Eq.(\ref{cgle}) with $c_1=c_2=0$) there is no long range order in the
system, so that there is no proper phase transition in  the thermodynamic 
$L\rightarrow\infty$ limit. Despite this, for large but finite  sizes and 
long but finite times, sharp transitions are observed and critical 
exponents and scaling functions can be consistently introduced 
\cite{emilio93}. This example should make clear that even in the case that
the PT-DT transition would not exist in the thermodynamic limit, its
characterization in large  finite systems is justified. 
The development of phase slips from PT waves  of high enough $\nu_i$ can be 
viewed as
a  kind of Eckhaus-like instability for turbulent waves, whereas the usual 
Eckhaus instability \cite{janiaud1} appears for regular waves.   This
similarity was one of the main motivations for the kind of  analysis that
follows.    
~

\begin{figure} 
\begin{center}
\epsfysize=55mm
\vspace{55mm}
{\epsffile{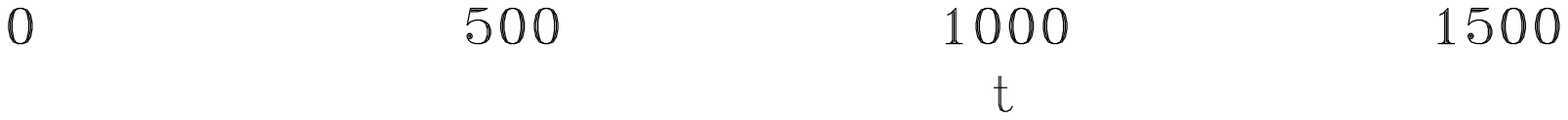}}
\vspace{-55mm}
\end{center}
\caption{\label{fig3}  a) Temporal evolution  of $\bar\nu (t)$ for four 
different initial
winding numbers $\nu_i = 25$ (solid), $20$(dotted), $15$ (dashed) $10$
(dashed-dotted). $c_{1} = 2.1, c_{2} = -0.83$ (PT regime). b) Winding number 
standard deviation $\sigma$.}
\end{figure}

\begin{figure} 
\begin{center}
\epsfysize=55mm
\vspace{55mm}
{\epsffile{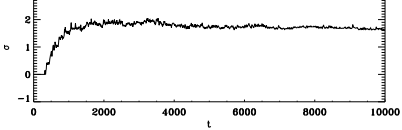}}
\vspace{-55mm}
\end{center}
\caption{\label{fig4} a) Temporal evolution  of $\bar\nu (t)$ for an initial 
winding number of $\nu_i = 4$ in the bichaos regime. 
$c_{1} = 1.6, 
c_{2} = -1.0$. b) Winding number standard deviation $\sigma$. }
\end{figure}

For each point in parameter space and initial winding number considered, we
have averaged over 50 independent random realizations of the  white Gaussian
perturbation added to the initial wave. Figs. \ref{fig3}a and \ref{fig3}b
show  the temporal evolution  of this average $\bar\nu(t)$ and its variance
$\sigma$ for $c_1=2.1$ and  $c_2=-0.83$. Four values of the initial  winding
number ($\nu_i= 10, 15, 20, 25$) are shown. Typically, the curve
$\bar\nu(t)$ decays from $\nu_i$ to a final winding number $\nu_f$. 
\pagebreak
\ \\
\vspace{-1cm}
~
The variance displays the behavior typical of a decay from a unstable state 
\cite{arecchi90}, namely a pronounced maximum at the time of
fastest variation of $\bar\nu(t)$. The final value of $\sigma$ gives the
dispersion in the final values of the winding numbers. Although the
behavior shown in Fig. \ref{fig3} is very similar to the observed in
\cite{emilio93} for a stochastic relaxational case, the scaling laws
found there do not apply here. The main qualitative difference is that in a
range of $\nu_i$ the sign of the average final $\bar\nu$ is here  opposite 
to the
initial one. In addition for some of the initial winding numbers (i.e.
$\nu_i=20$ in Fig. \ref{fig3}) $\bar\nu(t)$ is not monotonously
decaying, showing a small recovery after the fast decrease in $\bar\nu$.
These features are also observed for other values of $[c_1,c_2]$, so that   
figure \ref{fig3} is typical for $[c_1,c_2]$ in the PT region of Fig.
\ref{fig1}. For comparison  we show  $\bar\nu(t)$ and its
variance in Fig. \ref{fig4} for the point $c_1=1.6$  and $c_2=-1.0$, in the 
``bichaos" region.
The main difference is the existence of fast fluctuations in $\bar\nu$ and
$\sigma$. They are
related to the characteristic dynamics of the bichaos regime:  The final 
state 
depends on the initial conditions and it can correspond to PT, DT or even
coexistence of both. In the 50  realizations performed all  these
possibilities were found. When DT appears, there are big fluctuations
of the winding number  around $\nu=0$ that produce the wiggling on the
averaged value. More than  50 realizations should be performed to
smooth out such big fluctuations.   
~

\begin{figure} 
\begin{center}
\vspace{55mm}
\mbox{\subfigure{
\epsfysize=55mm
{\epsffile{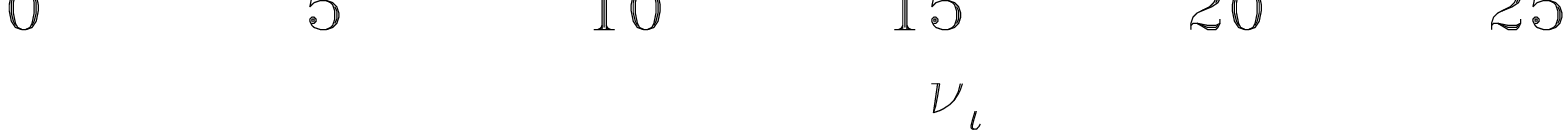}}
}\quad
\hspace{15mm}
\subfigure{
\epsfysize=55mm
{\epsffile{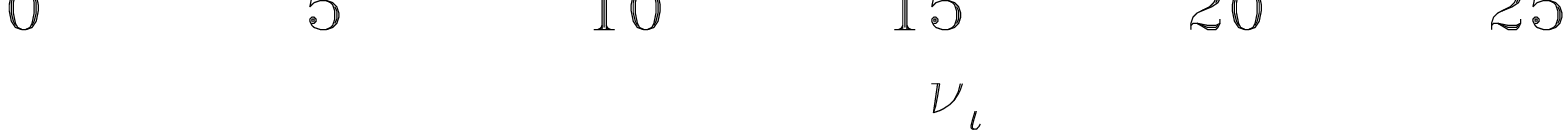}}
}\quad
}
\vspace{-55mm}
\end{center}
\caption{\label{fig5} a) Inverse of the characteristic time for winding number 
relaxation as a function of the initial 
winding number. The value of $c_{1}$ is fixed ($c_1=2.1$) and  $
c_{2}$ varies from near the BFN line ($c_2 \simeq -0.48$) to the $L_1$ line 
($c_2 \approx -0.9$). Different symbols correspond to $c_2= -0.6$ ($+$), 
$c_2 =-0.7$ ($\ast$), 
$c_2 = -0.75$ ($\Diamond$), $c_2 = -0.8$ ($\triangle$), $c_2= - 0.83$
 ($\Box$). The inset shows the critical winding number  ($\nu_c$) as a 
 function of  $ c_{2}$. b) Idem but the value of $c_{2}$ is fixed 
 ($c_2=-0.83$) and  $
c_{1}$ varies from near the BFN line ($c_1 \simeq 1.33$) to 
$c_1 = 2.5$ . Different symbols correspond to $c_1= 1.6$ ($+$), 
$c_1 =1.8$ ($\ast$), 
$c_1 = 1.96$ ($\Diamond$), $c_1 = 2.1$ ($\triangle$), $c_1= 2.3$ 
($\Box$), $c_1= 2.5$ ($\times$). The inset shows the critical winding number  
($\nu_c$) as a function of  $ c_{1}$}
\end{figure}

Returning to the PT parameter regime (Fig. \ref{fig3}) the decay of the
initial state  is seen to take place during a characteristic time  which
depends on $\nu_i$. We quantify this time   $\tau$ as  the time for which half
of the jump in $\nu$ is attained.  $\tau$ increases as $\nu_i$ decreases, and
there is a critical value of $\nu_i$, $\nu_c$,  such that no decay is observed
for $\nu_i<\nu_c$. Then $\tau$ diverges (critical slowing down) when 
$\nu_i$ approaches $\nu_c$ from above. This gives a sensible procedure to
determine $\nu_c$: Figs. \ref{fig5}a and \ref{fig5}b show $1/\tau$ as a 
function of $\nu_i$.  In Fig. \ref{fig5}a, $c_1$ is fixed and the different
symbols correspond to  different values of $c_2$. In Fig. \ref{fig5}b,
$c_2$ is fixed and the symbols correspond to different values of $c_1$. The
values of $\nu_c$ have been  estimated by extrapolating to $1/\tau = 0$ a
linear fit to the points of smallest $\nu_i$ in each  sequence. Motivated by
\cite{emilio93} we have tried to fit the divergence  of $\tau$ with
nontrivial critical exponents, but we have found no significant
improvement over the simpler linear fit. The values of $\nu_c$ so obtained 
are plotted in the insets of Figs. \ref{fig5}a and \ref{fig5}b. The range of
conserved winding numbers $[-\nu_c,\nu_c]$ is analogous to the Eckhaus
range of stable wavenumbers when working below the BFN line.  $\nu_c$ can
also be obtained by  directly determining the value of $\nu_i$ below which
$\nu(t)$ does not change in any of the realizations. This method can only
give integer values of $\nu_c$ whereas the method based on $\tau$ gives a
real number which is preferable when looking for continuous dependences
of $\nu_c$ on system parameters. The two methods however give consistent
results within the discretization indeterminacy.  
~

The insets of Fig. \ref{fig5}a and Fig. \ref{fig5}b indicate a clear
decrease in $\nu_c$  as the values of $c_1$ and $c_2$ approach the $L_1$ line. In
fact we know that  $\nu_c$ should be zero to the left of $L_1$, since no wave
maintains its winding number constant there. This lead us to a sensible
method for determining the  position of line $L_1$ \cite{montagne96b},
alternative to the one based in the density of defects used in
\cite{chate1}. It consists in extrapolating the behavior of $\nu_c$ to
$\nu_c=0$. A simple linear fit has been used. The same method to determine the 
line $L_1$ has been 
independently introduced in \cite{torcini196,torcini296}. The coefficients of
the linear fit are
not universal: they depend of the  particular path by which the line $L_1$ is
approached. With this method the line $L_1$ is determined as the line at 
which the range
of conserved winding numbers $[-\nu_c,\nu_c]$ shrinks to zero.  The
analogy with the Eckhaus instability of regular waves is again remarkable: 
in the same way as the range of Eckhaus-stable wavenumbers shrinks to zero  
when approaching the BFN line from below, the allowed $\nu$ range shrinks  to
zero when approaching the $L_1$ line from the right. The difference is that
below the BFN line the values of the wavenumber characterizes plane-wave
attractors, whereas above that line, $\nu$ characterizes
phase-turbulent waves.   In this picture, the transition line PT--DT
appears as the {\sl BFN line} associated to an Eckhaus-like instability for
phase turbulent waves. 
For the case of Fig. 5a the PT--DT  transition is located at $c_1=2.1,
c_2=-0.89\pm 0.02$, and  $c_1=2.60\pm 0.02, c_2=-0.83$ for the case of 
Fig. \ref{fig5}b.   The agreement with the position of the line as determined by
\cite{chate1,chate3}, where system sizes  similar to ours are used, is 
good.  For 
example for $c_1=2.1$ their value for $L_1$ is $c_2
=-0.92$.  The points marked as $+$ in Fig. \ref{fig1} 
correspond to  runs used to determine the position of the transition line
$L_1$ directly as the line at which defects appear in a long run even with
$\nu=0$. All these ways of determining $L_1$ give consistent results. Below 
the point $P$, $\nu_c$ goes
to zero when the parameters approach  the line $L_3$, not $L_1$, thus
confirming the known behavior that below point $P$ in Fig. 1 
the line separating
phase turbulence from defect turbulence when coming from the PT side is
actually $L_3$.    

The use of a linear fit to locate the line $L_1$ is questionable and more
complex fits have been tested. However, the simplest linear fit has been
found of enough quality for most of the the situations checked. Clearly some
theoretical guide is needed to suggest alternative functional forms for
$\nu_c(c_1,c_2)$. We notice that the analogous quantity below the BFN
line, the Eckhaus wavenumber limit, behaves as $q_E \sim \sqrt \epsilon$
for small $\epsilon$, being $\epsilon$  the difference between either
$c_1$ or $c_2$ and its value at the BFN line. From the insets in  Figs.
\ref{fig5}a or  \ref{fig5}b, this functional form is clearly less adequate
than the linear fit used.   

\begin{figure} 
\begin{center}
\epsfysize=55mm
\vspace{55mm}
{\epsffile{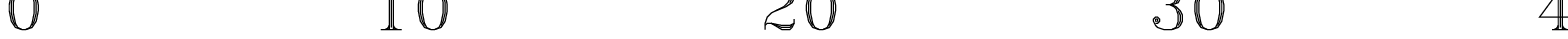}}
\vspace{-55mm}
\end{center}
\caption{\label{fig6} The final averaged winding number  ($\bar\nu_f$) as a 
 function of the initial one $\nu_i$. The initial condition is a TW with
 winding number $\nu_i$ for $c_1 = 2.1 \mbox{ and } c_2 = - 0.8$.
 The dashed line corresponds to the lowest of the two   Fourier modes of
fastest growth in  the linear regime as a function of $\nu_i$.}
\end{figure}

Another interesting point to study is the dependence of the final average 
winding number $\bar \nu_f$  on the initial  one $\nu_i$. Fig. \ref{fig6}
shows an    example using $c_1=2.1$ and $c_2=-0.8$. The behavior for other
values of the parameters is qualitatively similar.  $\bar \nu_f$ remains
equal to the initial value if  $\nu_i \leq 5$ during the whole simulation
time, so that $\nu_c \approx 5$, a value  consistent with the one obtained
from the divergence of $\tau$ and plotted in the inset of Fig.
\ref{fig5}a. For $\nu_i > \nu_c$, the final winding number is always
smaller than the initial one. By increasing  $\nu_i$ a minimum on $\bar
\nu_f$ is always observed, and then $\bar\nu_f$ tends to a constant value.
Figure \ref{fig6} also shows the winding number associated with  one of the 
two Fourier modes of fastest growth
obtained from the linear stability analysis of the initial traveling wave. 
The one shown is the lowest, the other one starts at $\bar\nu_f = 28$ and 
grows further up. Obviously they do not determine  the final state in a 
direct way. This 
is consistent with the observation mentioned above that the winding
number instability does not develop directly from the linear instability
of the traveling wave, but from a later nonlinear competition regime.   
~

\begin{figure} 
\begin{center}
\epsfysize=55mm
\vspace{55mm}
{\epsffile{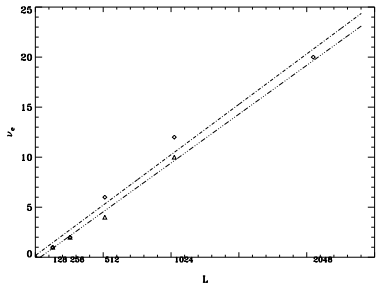}}
\vspace{-55mm}
\end{center}
\caption{\label{fig7} The critical winding number  ($\nu_c$) as a 
 function of the length $L$ of the system  is shown. Different symbols 
 correspond to $c_1= 2.1 \mbox{ and } c_2 = -0.8$ ($\triangle$), and 
 $c_1 = 1.96, \mbox{ and } c_2 = -0.83$ ($\Diamond$). The straight lines are
 linear fits to the two sets of data. }
\end{figure}

As stated in the introduction, a powerful way of distinguishing true phase
transitions from effective ones is the analysis of finite-size scaling
\cite{barber83}. We have tried to analyze size--effects from the point of
view of $\nu$ as an order parameter. In the DT state such kind of analysis was
performed in \cite{egolf}. Egolf showed that the distribution of the
values taken by the ever-changing winding number is a Gaussian function of 
width
proportional to $\sqrt{L}$. This is exactly the expected  behavior for
order parameters in disordered phases. In the thermodynamic limit the
intensive version of the order parameter, $\nu/L$, would tend to  zero so
that the disordered DT phase in the thermodynamic limit is  characterized by
a vanishing intensive  order parameter. For the PT states to be true
distinct phases,  the existence of a nonvanishing $\nu_c$ such that $\nu$ is
constant for  $|\nu|<\nu_c$ is not enough. The range of stable winding
numbers should also grow at least linearly with $L$ for this range to 
have any macroscopic significance.  The analysis of the growth of $\nu_c$
with system size has been performed in points 
$c_1 = 2.1, c_2 = -0.8 \mbox{ and } c_1 = 1.96, c_2 = -0.83$ of parameter 
space. $\nu_c$, determined as
explained before, is plotted in Fig. \ref{fig7} for several system sizes
for which the statistical sample of 50 runs was collected for each $\nu_i$. 
\pagebreak
\ \vspace{-1cm}
\\
~

There is a clear
increasing, close to linear, of $\nu_c$ as a function of $L$, thus indicating
that for this range of system sizes the range of allowed winding numbers is an
extensive quantity and then each $\nu$ is a good order parameter for
classifying well defined PT phases. It should be noted however that for the
larger system size for which extensive statistics was collected ($L=2048$)
data seem  to show a tendency towards saturation. Thus our study should be
considered as  not conclusive, and larger systems sizes need to be
considered.

\section{Different asymptotic states in the PT region}
\label{asymstat}

Typical configurations of the PT state of zero winding number consist of
pulses in the modulus $|A|$, acting as phase sinks, that travel and collide
rather irregularly on top of the $k=0$ unstable background wave (that is, a 
uniform oscillation)\cite{chate1,chate3}. The phase of these
configurations strongly resembles solutions of the 
Kuramoto-Shivashinsky (KS) equation. Quantitative agreement has been
found between the phase of the $\nu=0$ PT states of the CGLE and solutions  of
the KS equation near the BFN line\cite{egolf195}.   

For states with $\nu \neq 0$ a typical state\cite{chate3} is the one in which 
an average speed (in a direction determined by
the sign of $\nu$) is added to the irregular motion of the pulses.  We have
found that in addition to these  configurations there are other  attractors 
in  the PT region of parameters. We have identified \cite{montagne96b} three 
basic types of
asymptotic states for  $\nu \neq 0$, which we describe below. Other states
can be described in terms of these basic ones. Except when explicitly stated, 
all the configurations described in  this Section have been obtained by running
for long times Eq. (\ref{cgle}) with the initial conditions described
before, that is small random Gaussian noise added to an unstable traveling 
wave. The winding number of these final states is constant
and is reached after a transient period in which the winding number  might
have changed.  

\begin{figure}[H] 
\begin{center}
\epsfysize=120mm
\epsffile{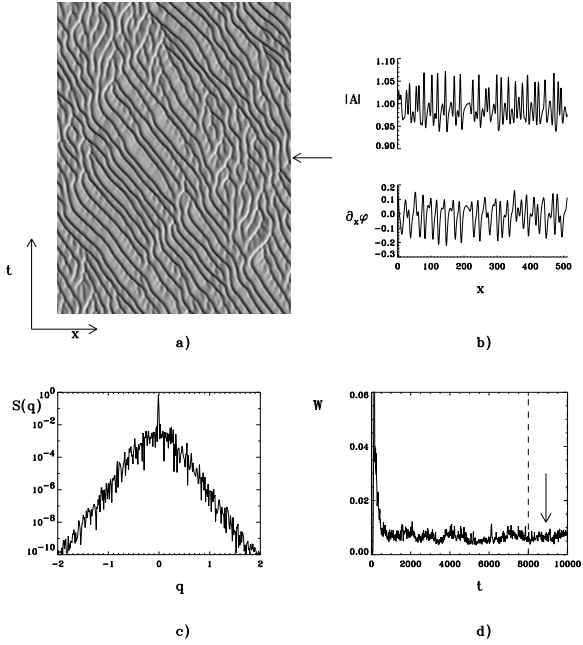}
\end{center}
\caption{\label{fig8} a) Spatiotemporal evolution of $\partial_x\varphi(x,t)$. 
The lighter grey correspond to the maximum value of $\partial_x\varphi(x,t)$ 
and darker to 
the minimum.  Last  $2000$ time units of a run $10^4$ time units long  
for a {\sl riding PT} state at $c_1 = 2.1$ , and $c_2 = - 0.83$. 
The  initial condition was a TW with  $\nu_i = 20$ that decayed to 
$\nu_f = -1$ after a short time.
b) A snapshot of   $\mid A(x,t) \mid  \mbox{ and }  \partial_x\varphi(x,t)$ as 
a function of $x$ for $t = 8900$ which 
is indicated by an arrow in a) and d).  
c) Spatial power spectrum $S(q)$ as a function of wavenumber 
at the same time $t = 8900$.
d)The time evolution of the quantity $W$ defined in the text. The dashed line
indicates the initial time for picture a).
 }
\end{figure}

\begin{figure} 
\begin{center}
\epsfysize=120mm
\epsffile{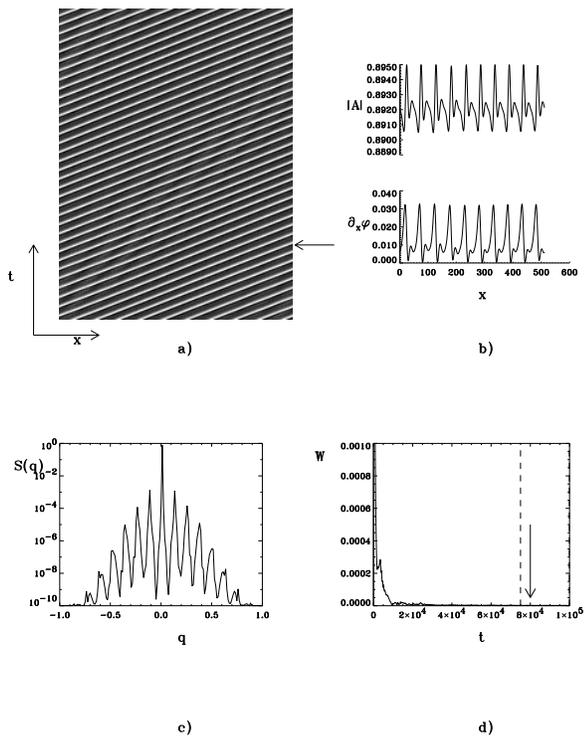}
\end{center}
\caption{\label{fig9} Same as in Fig. \protect\ref{fig8} but for last $35000$ 
time units of a run $10^5$  time units long for a quasiperiodic state. The
initial condition is random noise with an amplitude of $0.05$. 
$c_1 = 2.0$ , and $c_2 = -0.8$. b) and c) correspond to a time $t = 8 \times
10^4$.  }
\end{figure}

Figs. \ref{fig8}, \ref{fig9} and \ref{fig11} show examples of  the basic
states that we call  {\sl riding PT} (Fig. \ref{fig8}), {\sl quasiperiodic
states} (Fig. \ref{fig9}) and  {\sl frozen turbulence} (Fig.
\ref{fig11}).  For each figure: Panel (a) corresponds to a grey  scale
space--time plot of
$\partial_x\varphi(x,t)$. Panel (b) shows the value of this quantity and
the modulus of the field $(\mid A \mid )$ as a function of position
at the time indicated  by an arrow in panels (a)  and (d). 
Panel (c) shows the spatial power spectrum $S(q,t)$ of $A(x,t)$ for the same 
time. Finally, panel  (d) shows the quantity 
$W = \int \mid \partial_t S(q,t) \mid dq$, which is a global measure of the 
temporal change in the spatial power spectrum.   

{\sl Riding PT}. This state (see Fig. \ref{fig8}) is the most familiar one  
\cite{chate3}: wiggling pulses in the gradient  of the phase with a
systematic drift in a direction determined by $\nu$.  The modulus of the
field consists of a disordered spatial sequence of small pulses and shocks,
with $A(x,t)$ always far from zero. The spatial power spectrum $S(q)$ has a peak
corresponding to the global wave number $\bar k$ (associated in this case with 
$\nu = -1 $, so that $\bar k =2\pi\nu /L= -0.012$) and
a broad background associated with the turbulent motion ``riding" on the
traveling wave.  The time evolution of  $W$ shows a decay 
towards a fluctuating non-zero
value, indicating that the power  spectrum is continuously changing in time
as corresponds to the turbulent state  reached by the system. 
~

{\sl Quasiperiodic states}. These states (an example is shown in Fig.
\ref{fig9})  can be described as the motion of equidistant pulses in the
gradient of the phase that  travel at constant speed on top of the background
wave. The fact that the  periodicity of the pulses and that of the supporting
wave are not the same justify the name of  {\sl quasiperiodic}. We show  
later that these states  correspond to the ones described in Ref.
\cite{janiaud1}.  In Fig. \ref{fig9}a, the modulus $\mid A \mid$ and the
gradient of the phase clearly exhibit  uniformly traveling pulses.
The spatial power spectrum  $S(q)$  (Fig. \ref{fig9}c) clearly shows the
quasiperiodic nature of this state: a central peak, corresponding to the
dominant traveling wave, with equally spaced peaks surrounding it, showing the
periodicity of the pulses. The peaks are not  sharp because this
configuration has been obtained from a random
perturbation. The decrease of $W$  in Fig. \ref{fig9}d indicates that the
peaks are narrowing. Its asymptotic approach to zero indicates that the
amplitudes of the main modes reach a steady value and $S(q)$ becomes  
time independent.    
~

\begin{figure} 
\begin{center}
\epsfysize=50mm
\vspace{50mm}{\epsffile{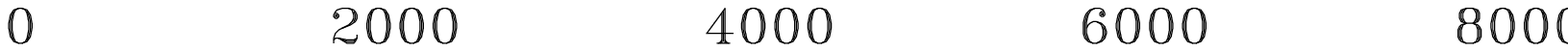}}
\vspace{-50mm}\end{center}
\caption{\label{fig10} a)The time evolution of $W$ 
for a quasiperiodic state. The initial condition is
a TW sinusoidally perturbed for $c_1 =2.1 \mbox{ and } c_2 = -0.6$ . 
In the inset the spatial power spectrum $S(q)$ as a function of wavenumber 
at the time $t = 8900$ indicated by an arrow in the main picture.}
\end{figure}

\begin{figure} 
\begin{center}
\epsfysize=120mm
\epsffile{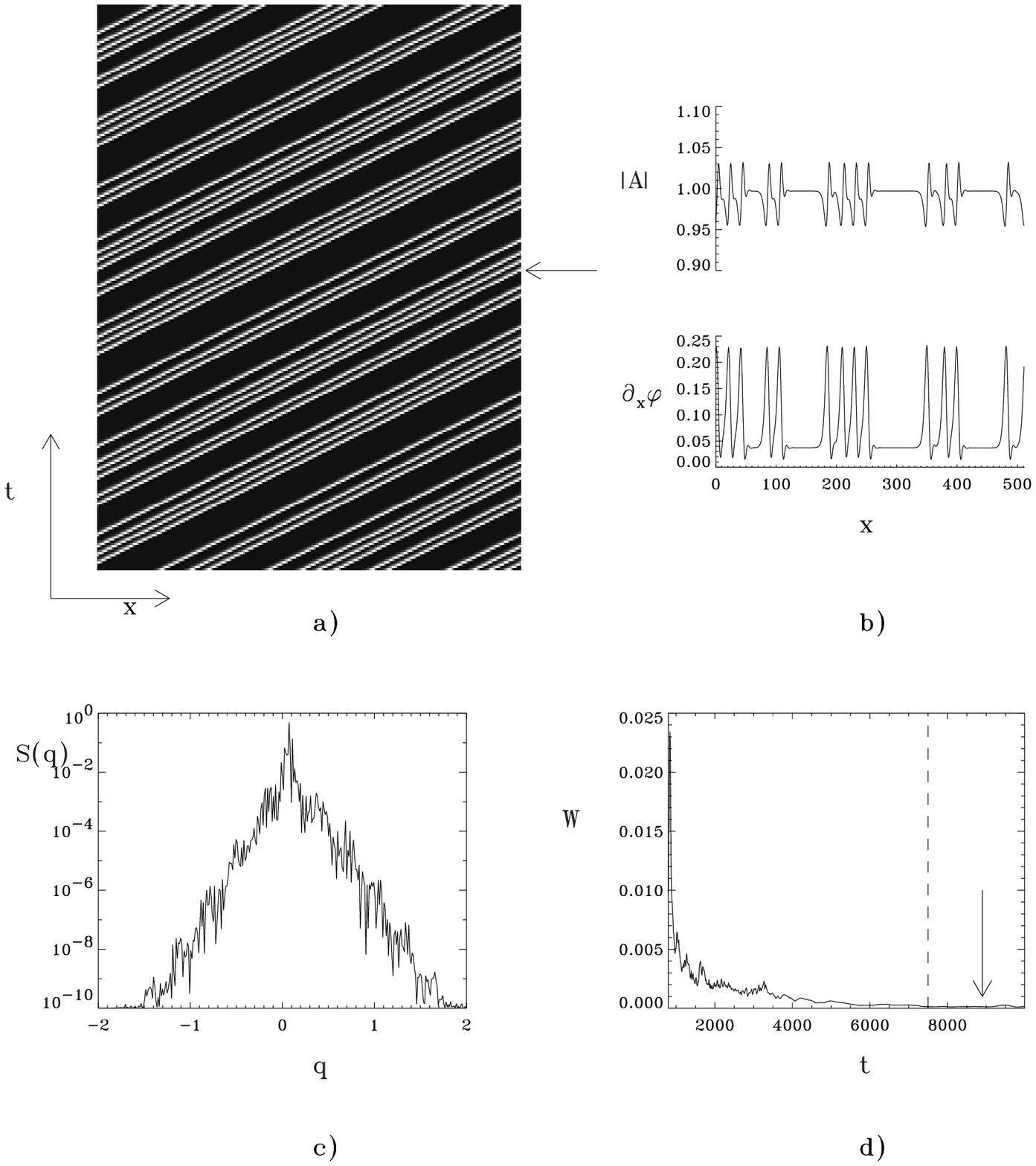}
\end{center}
\vspace{-05mm}
\caption{\label{fig11} Same as in Fig. \protect\ref{fig9} but for last $2500$
 time units of a run $10^4$  time units long for a frozen 
turbulence  state. The initial condition was a TW 
of $\nu_i = 12$ that decayed to $\nu_f = 6$ after a short time. $c_1 = 1.75$  
and $c_2 = -0.8$. The time of b) and c) is $t=8900$,  indicated by
an arrow as in previous figures.}
\end{figure}

More perfect quasiperiodic configurations can be obtained from initial
configurations that are already quasiperiodic. Figure \ref{fig10} shows the
quantity $W$ for a state generated at $c_1 = 2.1 \mbox{ and }
c_2 =-0.6$ from an initial traveling wave with a sinusoidal
perturbation. The initial traveling wave had $\nu_i = 18$ and the winding
number of the sinusoidal perturbation was $\nu = 22$. The travelling wave
decayed to a state with $\nu_f=10$ of the quasiperiodic type, cleaner than
before. The 
spatial  power spectrum (shown in the inset at the time indicated by an 
arrow in the main picture) shows the typical characteristics of a 
quasiperiodic state.
\pagebreak
\ \vspace{-1cm}
\\
~

{\sl Frozen turbulence}. This state (see Fig. \ref{fig11}) was first described
in \cite{montagne96b}. It consists of pulses in $\partial_x\varphi$ traveling 
at constant speed on a traveling wave background, as in the quasiperiodic case, 
but now the pulses are not equidistant from each other (see Fig. \ref{fig11}b).
The power spectrum at a given time is quite different from 
the one of  a quasiperiodic state. It is similar, instead, to the power
spectrum obtained in the {\sl riding PT} state: $S(q)$ is  a broad 
spectrum in the sense that the inverse of the width, which gives a measure
of the correlation length, is small compared with the system size.
Here however $W$ relaxes to zero, so that the power 
spectrum finally stops changing (thereby the name { \sl frozen}). 
~
This behavior is an
indicator of the fact \cite{montagne96b}, obvious from Fig. \ref{fig11}, that 
the pattern approaches a
state of rigid motion for the modulation in modulus and gradient of the phase 
of the unstable background plane
wave. That is, the field $A(x,t)$ is of the form:  
\BE
\label{uniftrans0}
A(x,t)=g(x-vt) e^{i\left( kx-\omega_k t + \alpha(t) \right)}
\EE
where $g$ is a uniformly translating complex modulation factor.   
It is easy to see that configurations of the form (\ref{uniftrans0})  have a
time-independent spatial power spectrum. Torcini \cite{torcini196} 
noticed in addition that the function $\alpha(t)$ is linear in $t$ so that the
solutions are in fact of the form 
\BE
\label{uniftrans}
A(x,t)=f(x-vt) e^{i(kx-\omega t)}
\EE
where again $f(x-vt)$ is a complex valued function and $\omega$ can differ from
$\omega_k$. $f$ and $g$ differ only in a constant phase. The envelopes  
$g(x-vt)$ or $f(x-vt)$ turn out to be rather
irregular functions in the present {\sl frozen turbulence} case, whereas they
are periodic in the quasiperiodic configurations discussed above. 

\begin{figure} 
\begin{center}
\epsfysize=120mm
\epsffile{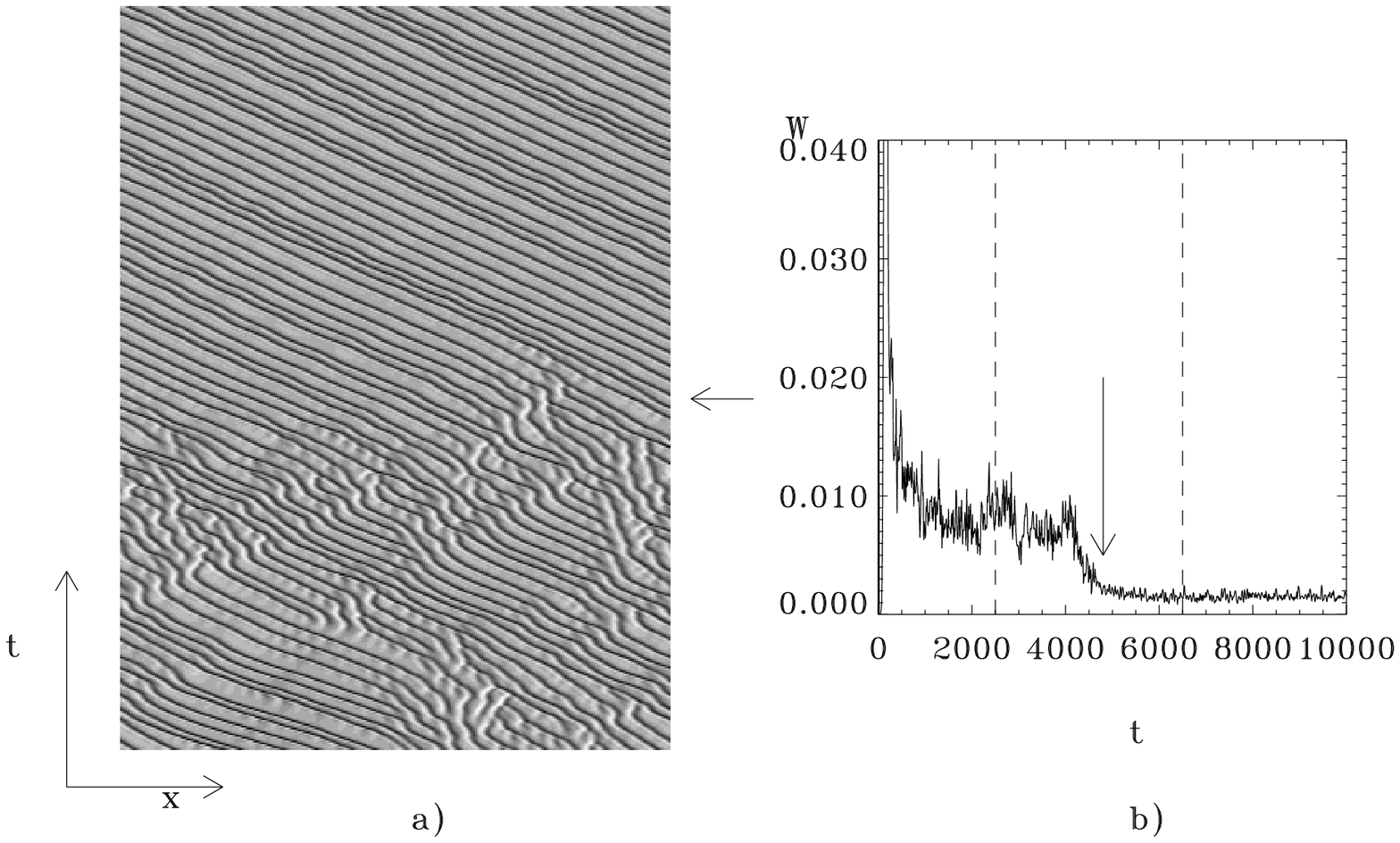}
\end{center}
\vspace{-50mm}
\caption{\label{fig12}a) Spatiotemporal evolution of $\partial_x\varphi(x,t)$ 
for  
a riding turbulence state that decays onto another one. 
$c_1 = 2.5$, $c_2 = -0.75$. The initial condition was 
a TW of $\nu_i = 20$ that decayed to $\nu_f = -2$ in a short time. 
b) Time evolution of $W$. The dashed lines indicate the time interval shown 
in a) (from $t_1 = 2500$ to $t_2 =6500$ of a run $10^5$ time units long). 
The arrow indicates the transition from one of the riding turbulence regimes 
to the other one. }
\end{figure}

\begin{figure} 
\begin{center}
\epsfysize=70mm
\epsffile{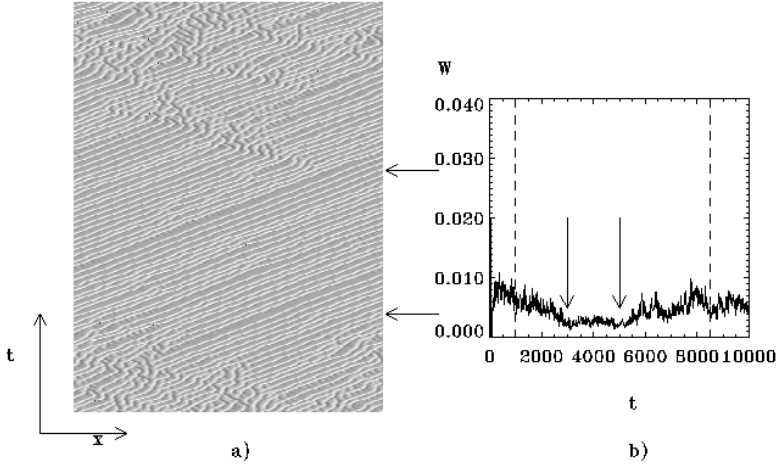}
\end{center}
\vspace{-0mm}
\caption{\label{fig13} a) Spatiotemporal evolution of $\partial_x\varphi(x,t)$ 
showing intermittency between riding turbulence states. $c_1 = 2.1$,   
$c_2 = -0.83$. The initial condition 
is a TW of $\nu_i = 1$ that did not change.  
b)Time evolution of $W$. The dashed lines indicate the time interval shown in 
a) (from $t_1 = 1000$ to $t_2 =8500$ of a run $10^4$ time units long). The
arrows indicate the end of a riding turbulence regime and the beginning of 
another one.
}
\end{figure}

After presenting the basic states, we continue addressing some
interesting mixed states  that can be described in terms of them. Most of the
configurations ending up in the frozen turbulence or in the quasiperiodic states
have long time transients of the riding turbulence type. Only at long times a 
decay
to rigid propagation occurs. There are cases in which a different type of
decay happens. For example  Fig. \ref{fig12} shows a case in which the system 
jumps from a very strong  
riding turbulence regime to another state, also of the riding turbulence 
type, but much more regular. The quantity $W$, shown in 
Fig. \ref{fig12}b, turns out to be a valuable tool in distinguishing the
different regimes: a superficial look at Fig. \ref{fig12}a could be easily 
misunderstood as
indicating the approach of the system towards a frozen turbulence state, but
the lack of decay towards zero of $W$ identifies the final state as 
riding turbulence. The arrows indicate the jump to the second state. Fig.
\ref{fig13} shows a state characterized by a recurrence 
between two different riding turbulence states, showing a kind of temporal 
intermittency. 
\vspace{-10mm}
~

\vspace{-0mm}
~

\begin{figure} 
\begin{center}
\epsfysize=100mm
\epsffile{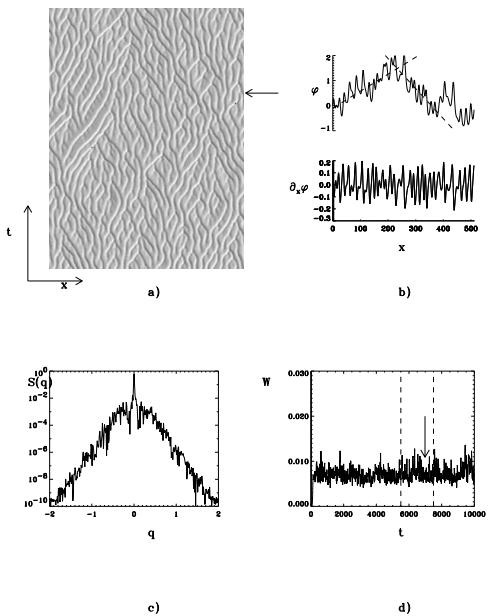}
\end{center}
\caption{\label{fig14} a) Spatiotemporal evolution of $\partial_x\varphi(x,t)$.
 The time interval corresponds to  $5500 \mbox{ to } 7500$ time units of a 
run $10^4$ time units long  for a {\sl riding PT} state at $c_1 = 2.1$  
and $c_2 = - 0.83$. The  initial condition was a TW with  $\nu_i = 0$ that 
did not change. 
b) Snapshots of $\varphi(x,t)  \mbox{ and }  \partial_x\varphi(x,t)$ as a
function of $x$ at the time $t = 6980$ indicated by an arrow in a) and d).  
The dashed lines in the graph of $\varphi(x,t)$ indicate average slopes, 
that is ``local" wavenumbers. 
c) Spatial power spectrum $S(q)$ as a function of wavenumber 
at the same time $t = 6980$. d)Time evolution of $W$. Dashed 
lines indicate the time interval of picture a).  }
\end{figure}

Finally Fig. \ref{fig14} shows a riding turbulence state with zero 
winding number. This is not 
however a typical configuration, since usually for $\nu=0$ there is no
preferred direction for the pulses to drift, whereas the figure shows that 
in fact there is a local drift at some places of the system. It turns out that 
this state can be 
understood as composed by two domains of different local winding number: 
$\nu = 1$ and $\nu = -1$, so that globally $\nu=0$. The 
pulses travel either in one direction or in the other depending of the region
of the system in which they are. In Fig. \ref{fig14}b a snapshot of the 
gradient of the phase 
$\partial_x\varphi(x,t)$ and
the phase itself $\varphi(x,t)$ is shown. Lines showing the average trend 
in the phase are plotted over the phase, clearly identifying the two regions in
the system. This coexistence of the different
basic states at different places of space, or at different times as in 
Fig. \ref{fig13}, was already mentioned in 
\cite{chate2} where it was argued to give rise to a kind of spatio-temporal
intermittent behavior. 
~

Given the large variety of configurations that are observed, and the very long
transients before a jump from one state to another occurs, it would be 
difficult to 
conclude from numerical evidence alone that the three kinds of states
considered as {\sl basic} above are true asymptotic states. Some analytical
insight would be desirable to be sure that these three states are attractors of
the dynamics. The next Section is devoted to provide such analytical
justification.  

\section{Asymptotic states in terms of the phase dynamics}
\label{kawa}

The question on whether it is possible or not to describe the PT 
regime of the CGLE from a closed equation for the phase alone has been posed
by several
authors\cite{chate8,egolf195,chate9,sakaguchi1,grinstein96}. A phase
equation is obtained by considering a long wavelength perturbation of a 
plane-wave solution in the CGLE
(\ref{cgle}).  It is clear that this phase equation will only describe phase
dynamics close to the homogeneous plane-wave (that is the one with $\nu = 0$) 
if the perturbation is made around the
spatially--homogeneous solution. In order to get a
description of PT at $\nu \neq 0$ the expansion 
should be done for a perturbation on a traveling wave solution
with wavenumber ($k$) different from zero,
 \begin{equation}
A = (\sqrt{1 - k^2} \, + a(x,t)) 
e^{i\left( k x + \phi(x,t) \right)} \; ,
\label{slowperk}
\end{equation}

Here $k$ is taken as $k = \frac{2 \pi}{L} \nu$. If $A$ satisfies 
periodic boundary conditions the same conditions apply to $\phi$ because
any global phase winding is included in $k$ (the total phase is 
$\varphi = k x + \phi$). 
From general symmetry arguments the general phase equation for $k \neq 0$
should read, up to fourth order in gradients:   

\begin{eqnarray}
\partial_t \phi &=& \Omega_0 - v_g \partial_{x} \phi - D_2 \partial_{x}^2 \phi
+D_{11}  (\partial_{x} \phi)^2 + D_3 \partial_{x}^3 \phi  + 
D_{12}(\partial_{x} \phi)(\partial_{x}^2 \phi) \nonumber \\ 
&-&  D_4 \partial_{x}^4 \phi + 
D_{13}(\partial_{x} \phi)(\partial_{x}^3 \phi) + 
D_{22} \left(\partial_x^2 \phi \right)^2 + 
D_{112}(\partial_{x} \phi)^2(\partial_{x}^2 \phi) + \ldots \; .
\label{exkseq}
\end{eqnarray}
When $v_g = D_3 = D_{12} = D_{13}= D_{22}= D_{112}=0$, Eq. (\ref{exkseq})
reduces to the Kuramoto--Sivashinsky (KS) equation \cite{kuramoto78,siva77} 
that is the lowest order nonlinear phase equation for the case $k = 0$. For $k
\neq 0$, Eq. (\ref{exkseq}) was systematically derived up to third order in 
gradients in \cite{legath}.  An easy way of obtaining the values of all the
coefficients in (\ref{exkseq}) was discussed in \cite{kuramoto84}: First,
$\Omega_0$ is related to the frequency of the plane-wave solutions: 
\BE \Omega_0 = - \omega_k = -c_2 - (c_1 -c_2) k^2 \EE
Second, the
linear terms can be obtained from the eigenvalue $\lambda(k,q)$ 
corresponding to the phase-like branch in the linear stability analysis of the
wave of wavenumber $k$ with respect to perturbations of wavenumber
 $q$\cite{janiaud1,hohenbergsaarloos,legath,montagne96e}: 
\BE
\lambda(k,q)= -i v_g q + D_2 q^2 - iD_3 q^3 -D_4 q^4 + {\cal O} (q^5)
\label{eigen}
\EE
\noindent with 
\begin{eqnarray}
v_g &=&  2 k (c_1 -c_2) \label{vg}\\
D_2 &=& -(1 + c_1 c_2) + {2 k^2 (1+c_2^2) \over 1-k^2 }  \label{D2}\\
D_3 &=& {2 k ( 1 + c_2^2) \left[ -c_1 +( c_1+2 c_2 ) k^2 \right] \over 
\left( 1-k^2 \right)^2 }  \label{D3}\\
D_4 &=& \frac{1}{2 \left( 1-k^2 \right)^3} 
\left\{ c_1^2 (1+c_2^2) -2 k^2 (1+c_2^2) (c_1^2 + 6 c_1 c_2)  \nonumber \right.\\
&+& k^4 \left. \left[ 4 + (1+c_2^2)  
 (c_1^2 +12 c_1 c_2) + c_2^2 (24+20c_2^2)  \right] \right\} \; .
\label{D4}
\end{eqnarray}
Third, the nonlinear terms can be obtained from the following consistency
relationship: If (\ref{slowperk}) is 
an exact solution
of the CGLE, then $\phi(x,t)$ satisfies the phase equation with coefficients
depending on $k$. In addition if  
$(\sqrt{1-k_1^2}+a_1(x,t)) e^{i(k_1x+\phi_1(x,t))}$ 
is another exact solution of the CGLE, then $\phi_1(x,t)$ satisfies a similar
phase equation but with coefficients depending on $k_1$ instead of $k$. But
this solution can be written as $(\sqrt{1-k_1^2}+a_1(x,t))
e^{i(kx+(k_1-k)x+\phi_1(x,t))}$ so that $(k_1-k)x+\phi_1(x,t)$ is also
solution of the phase equation with coefficients depending on $k$ (with
different boundary conditions). By combining the two equations satisfied by
$\phi_1$ and expanding the coefficients depending on $k_1$ as a power series
around $k$ (assuming $k_1-k$ small) the following relationships between linear 
and nonlinear terms are obtained: 
\BE
D_{11}= - {1\over 2}{\partial v_g \over \partial k}\ ,\ \ \ 
D_{12}= - {\partial D_2 \over \partial k} \ ,\ \ \ 
D_{13}={\partial D_3 \over \partial k}\ ,
\ \ \ D_{112}=-{1\over 2}{\partial^2 D_2 \over \partial k^2}
\EE
So that 
\begin{eqnarray}
D_{11} &=& c_2-c_1 \label{D11}\\
D_{12} &=& - {4 k (1+c_2^2) \over \left(1-k^2 \right)^2} \label{D12}\\
D_{13} &=& {2(1+c_2^2) \over\left(1-k^2 \right)^3} 
\left[-c_1+6c_2k^2+(2c_2+c_1)k^4 \right]   \label{D13}\\
D_{112} &=& -{2(1+c_2^2)(3k^2+1) \over \left(1-k^2 \right)^3}
\label{D112}
\end{eqnarray}
The coefficient $D_{22}$ is only obtained following the method to higher order
in 
$(k_1 - k)$. The coefficients up to third order in gradients can be found also 
in \cite{legath} and approximate expressions for them are given in 
\cite{janiaud1}. 

The traveling wave of wavenumber $k$ becomes unstable when the coefficient 
$D_2$ becomes positive. One expects that the first terms in the gradient 
expansion 
(\ref{exkseq}) give a good description of the phase dynamics in the weakly 
nonlinear regime, that is $D_2$ positive but small (note that for a given  $k\neq 0$
this includes part of the region below the BFN line in Fig 1). The arguments 
presented in
\cite{kuramoto84} imply that the relative importance of the
different terms in a multiple scale expansion in which $D_2$ is the small 
parameter can be established by considering 
$\phi \sim \partial_x \sim D_2^{1/2}$. Then the dominant terms close to the
instability of wave $k$ are the ones containing $\Omega_0$ and $v_g$. After
them, the terms with coefficients $D_3$ and $D_{11}$ are the most
relevant. Up to this order Eq. (\ref{exkseq} ) is   a Korteweg--de Vries 
equation
(KdV). The terms with $D_2$, $D_4$ and $D_{12}$ appear at the next
order. The importance of the terms in $D_2$ and $D_4$ for a qualitatively 
correct description of phase dynamics is obvious since they
control the stability properties of the wave of wavenumber $k$. The importance 
of the term with coefficient $D_{12}$ was stressed in \cite{janiaud1,nepo195}: 
if it is large
enough it can change the character of the bifurcation from supercritical to
subcritical. 

The detailed comparison of the reduced dynamics (\ref{exkseq}) with the 
complete CGLE phase dynamics is beyond the scope of the present paper. 
The aim of this Section is to  use  Eq.(\ref{exkseq}) just to get some
understanding of the asymptotic states presented in Section \ref{asymstat}. 
To this end we will use the detailed results available from the work of Chang 
et al. \cite{chang93}. These results are obtained for the so-called Kawahara
equation \cite{kawa83,kawa85,nepo195,chang93,kawa88} which is Eq.(\ref{exkseq}) 
with $D_{12}=D_{13}=D_{22}=D_{112}=0$. The term $D_{12}$, which according to 
Kuramoto
estimations \cite{kuramoto84} is of the same order for small $D_2$ as the terms
in $D_2$ and
$D_4$, will thus be neglected. It would be certainly necessary to consider the
modifications introduced by the term $D_{12}$ into the results of
\cite{chang93}. This will be briefly discussed at the end of this section. 
At this
point it is interesting to note that, to our knowledge, the only quantitative 
comparison of the phase dynamics with $k \neq 0$ obtained from a phase equation and from 
CGLE is \cite{torcini196,torcini296}. But the phase equation used in these
references is 
the one presented in \cite{sakaguchi1}, in which the nonlinear terms 
considered are only those with coefficients 
$D_{11}$ and $D_{13}$. In addition $D_{11}, D_{13}$, and the coefficients of 
the linear terms are considered only up
to first order in $k$. Despite these limitations, in particular the absence of the 
$D_{12}$ term, the phase
equation is found to reproduce well the phase dynamics of the CGLE, an
agreement that degrades when the term in  $D_{13}$ is suppressed
\cite{torcinipc}. Clearly further work is needed to establish firmly the 
relevance
of the different terms in (\ref{exkseq})\cite{blowup}. Our study will be 
restricted to
the situation of \cite{chang93} (that is $D_{12}=D_{22}=D_{13}=D_{112}=0$) since no 
study of comparable detail  for a more complete equation is available in the 
literature. 

The situation of interest here is the one in which the traveling waves are
unstable against a finite band of wavenumbers, so that $D_2,\ D_4>0$. 
Making the following changes of variables 
in (\ref{exkseq}) with 
$D_{12}=D_{22}=D_{13}=D_{112}=0$: 
\begin{eqnarray}
\chi &=& \sqrt{\frac{D_2}{D_4}}\left( x -v_g t\right) 
\; ,\nonumber \\
\tau&=&\frac{D_2^2}{D_4} t  
\; , \nonumber \\
u (\chi,\tau) &=& 
-\frac{D_{11}D_4^{1/2}}{2 D_2^{3/2}}\partial_x \phi(x,t) \; .
\label{scaling}
\end{eqnarray}
the  Kawahara equation\cite{kawa83,kawa85,nepo195,chang93,kawa88} 
is obtained
\begin{equation}
\partial_{\tau} u = - \partial^2_{\chi} u -4 u \partial_{\chi} u - \delta 
\partial^3_{\chi} u  - \partial^4_{\chi} u \; ,
\label{kawae}
\end{equation}
\noindent with 
\begin{equation}
\delta = - \frac{D_3}{\sqrt{D_2 D_4}}
\label{delta}
\end{equation}
Since $\phi$ is periodic in $x$, $u(\chi,t)$ is periodic in $\chi$. 
In addition  $\int_0^L u(\chi,\tau) d\chi=0$. 
To have some intuition on the meaning of the parameter $\delta$, its expansion at small
$k$ reads
\begin{equation}
\delta \approx 2 \sqrt{2}\, k \, 
\sign(c_1) \sqrt{\frac{1 + c_2^2}{\mid 1 + c_1 c_2 \mid}} + 
{\cal O}(k^3)\; .
\label{deltaex}
\end{equation}
It should be noted that $\delta$ does not diverge at the BFN 
line, as the expansion 
(\ref{deltaex}) seems to  suggest, but below it. From Eq. (\ref{delta}) it is 
clear that $\delta$ diverges  where $D_2$ vanishes indicating that 
the corresponding traveling wave of wavenumber $k$ has become Eckhaus unstable. 

The Kawahara equation (\ref{kawae}) has been considered in the context of 
surface
waves on fluid films falling down an inclined or vertical plane \cite{chang94},
and also as a simple generalization of the KS or the KdV equations
\cite{kawa83,kawa85}. It has also been considered in the context of growth
shapes \cite{conrado94}. It reduces to KS for $\delta = 0$ 
(or equivalently for $k = 0$) when written in the original variable $\varphi$. 

Equation (\ref{kawae}) has periodic,
soliton-like, spatially-irregular, and spatio-temporally chaotic solutions. 
\cite{kawa83,kawa85,kawa88}. In fact, all of these solutions have been
analytically shown to exist \cite{chang93}. All of them except the
isolated soliton-like solution \cite{chang95n} are stable in some parameter 
regimes \cite{chang93}. These 
kinds of solutions should manifest themselves (provided the approximate phase
description holds) in the time evolution of the phase gradient $\partial_x 
\varphi$ ($=k + \partial \phi$)  of the
solutions of the CGLE (\ref{cgle}) in the PT regime. The analytical results in 
\cite{chang93} thus provide a firm basis for true existence of the numerically 
observed states described in Section \ref{asymstat}. 

\begin{figure} 
\begin{center}
\epsfysize=120mm
\epsffile{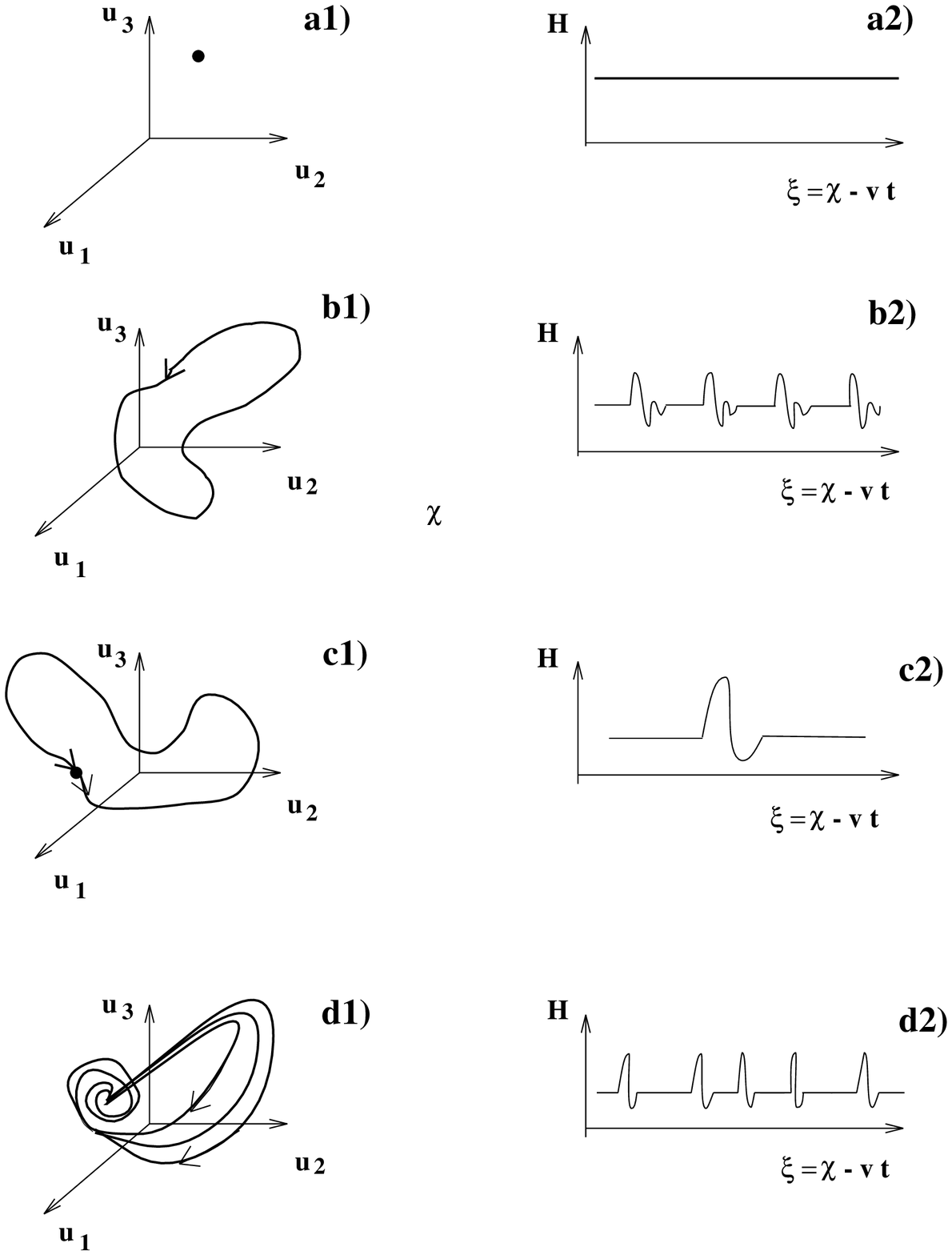}
\end{center}
\caption{\label{fig15} Schematic relationship between trajectories of 
the dynamical system
(\protect\ref{dynsys}) in phase space (left column) and solutions 
$u(\chi,\tau)=
H(\xi=\chi - v \tau)$ of equation (\protect\ref{kawae}) (right column). 
a1) Fixed point of (\protect\ref{dynsys}), a2) uniform solution of
 (\protect\ref{kawae}) (traveling wave in the CGLE (\ref{cgle})). 
b1) Periodic solution (limit cycle) of (\protect\ref{dynsys}), b2) periodic 
train solution of
(\protect\ref{kawae}) (quasiperiodic solution of the CGLE 
(\protect\ref{cgle})). 
c1) Homoclinic trajectory of (\protect\ref{dynsys}), c2)  single pulse of
 (\protect\ref{kawae}). 
d1) Chaotic trajectory of (\protect\ref{dynsys}), d2)  spatially 
irregular solution of (\protect\ref{kawae}) (frozen turbulence in the
CGLE). }
\end{figure}

The detailed bifurcation analysis in \cite{chang93}   also gives 
detailed predictions for the wound states of the CGLE, within the range of
validity of the phase description. We will reproduce here some of the 
results in \cite{chang93} and reinterpret them in terms of the gradient of 
the phase of CGLE
solutions. Our interest is centered in the rigidly moving train of pulses
( frozen turbulence and quasiperiodic states)
observed in several of the numerical simulations reported in 
Section \ref{asymstat}. They are of the form (\ref{uniftrans}), and 
because of (\ref{scaling}) we have
\BE
u(\chi,\tau) = H(\xi)\ ,
\EE
with $\xi = \chi - v \tau$, being $v$ the velocity of the train of pulses we 
want to describe in units of $\chi \mbox{ and } \tau$. The partial differential
equation (\ref{kawae}) is reduced to an ordinary differential equation
(ODE) for $H(\xi)$:  
\begin{equation}
H^{iv} + \delta H''' + H'' +4 H H' - v H' =0 \; .
\label{ode1}
\end{equation}
The primes denote differentiation with respect to $\xi$. After an 
integration: 
\begin{equation}
 H''' +\delta H'' + H' - v H + 2 H^2 = Q \; .
\label{ode2}
\end{equation}
$Q$ is fixed in a nontrivial way by the condition $\int H d\xi =0$ which follows
from our periodic boundary conditions. This third order
 ODE  can be rewritten as a three-dimensional dynamical system: 
\begin{eqnarray}
u'_1 &=& u_2 \nonumber \\
u'_2 &=& u_3 \nonumber \\
u'_3 &=& c u_1 - u_2 - \delta u_3 - 2 (u_1)^2 
\label{dynsys}
\end{eqnarray}

\noindent with 
\begin{eqnarray}
u_1(\xi) &=& H(\chi) - \frac{v}{4} + \sqrt{\frac{c^2}{16} + \frac{Q}{2}}
\nonumber \; , \\
c &=& \sqrt{8 Q + v^2} \; .
\label{uuno}
\end{eqnarray}
Different qualitative behaviors in phase space of the solutions of the 
dynamical system (\ref{dynsys}) are related to the shape of the solutions of
(\ref{ode1}) \cite{pumir83}. This is illustrated in Fig. \ref{fig15}.
~

  We stress 
that all the solutions 
of (\ref{ode1}) represent uniformly
translating solutions of (\ref{kawae}). No information is given on more 
complicated solutions of (\ref{kawae}). 
The left column of Fig. \ref{fig15} shows the possible trajectories of the dynamical 
system 
(\ref{dynsys}) while the right column shows the corresponding solution of 
(\ref{ode2}), or equivalently $u(\chi,\tau)=H(\xi=\chi - v \tau)$ in equation 
(\ref{kawae}). 
For a fixed point in (\ref{dynsys}) (fig. \ref{fig15}a1) we get a
homogeneous solution in (\ref{kawae}) (fig. \ref{fig15}a2) and (via
(\ref{scaling})) 
a traveling wave
solution in the CGLE (\ref{cgle}). For a periodic trajectory in
(\ref{dynsys})
(fig. \ref{fig15}b1) we get a train of periodic pulses in the solution of 
(\ref{kawae}) 
(fig. \ref{fig15}b2) and a quasiperiodic solution in CGLE (\ref{cgle}). An 
homoclinic
trajectory in (\ref{dynsys}) (fig. \ref{fig15}c1) corresponds to a single pulse 
solution
in (\ref{kawae}) (fig. \ref{fig15}c2). Finally for a chaotic trajectory in
(\ref{dynsys}) (fig. \ref{fig15}d1) we have an irregular solution $H(\xi)$ that 
corresponds
to a rigidly traveling spatially irregular solution of (\ref{kawae}) 
(fig. \ref{fig15}d2). The 
chaotic solutions of (\ref{dynsys}) are of the Shil'nikov type \cite{chang93}. 
This means that the disordered configurations $H(\xi)$ (and thus $u$ and 
$\partial_x\varphi$) consist on nearly identical pulses irregularly spaced. 
This corresponds to the state named {\sl frozen turbulence} for solutions of
the CGLE. 

The detailed analysis of \cite{chang93} is done on the one hand by following
the sequence of bifurcations of the state in which $H$ is a constant and of 
the state
in which $H$ is close to the KdV soliton (with adequate rescaling Eq.
(\ref{kawae}) reduces to the KdV in the limit $\delta \rightarrow \infty$). 
On the other hand the powerful global theorems of Shil'nikov and their
generalizations \cite{wiggins1,nayfeh1,kuznetsov,shilnikov} are used to 
establish the structure of the solutions of (\ref{dynsys}). The results 
of \cite{chang93} relevant to
our purposes can be summarized as follows (they can be read-off from 
figure 3 of Ref. \cite{chang93} ):   
\begin{enumerate}
\item Periodic solutions of (\ref{dynsys}) exist for all values of $\delta$
provided $\mid c \mid > \mid \delta \mid $. They are organized in a variety of
branches. Solutions in the same branch differ by their periodicity, and each
branch ends in a different kind of solitary-wave solution (infinite spatial
period). The shape of the different solitary wave solutions characterizes the
different branches.
\item For $\mid\delta\mid \gtsim 1.1$ only one of the branches of periodic 
solutions (the {\sl main branch}) remains. 
\item Chaotic solutions to (\ref{dynsys}) exist only for 
$\mid\delta\mid\lesssim 0.84$. 
\end{enumerate}
In addition Chang et al. \cite{chang93} obtained results also for the full 
equation (\ref{kawae}), without the restriction to rigidly traveling waves. 
Their numerical and analytical results can be summarized as 
\begin{enumerate}
\setcounter{enumi}{3}
\item Periodic solutions in the main branch with its wavenumber within a given
range are linearly stable for all $\delta$. A more precise determination of the
range of stable wavenumbers for large $\delta$ was performed in \cite{nepo195}.
\item In addition to the periodic solutions there are also spatio-temporal
chaotic attractors for all $\delta$.
\item If $\mid\delta\mid > 1.1$ only two of these strange attractors remain. 
For $\mid\delta\mid >3$ their basin of attraction seems to be much smaller 
than the one of the periodic solutions. 
\end{enumerate}

Expression
(\ref{delta}) with (\ref{vg})-(\ref{D112}) gives the relation 
between $\delta$ and the parameters of the CGLE.
$\mid\delta\mid=\infty$ corresponds in Fig. 1 to the line at which the wave of 
wavenumber $k$ becomes Eckhaus unstable. It is approximately parallel and below
the BFN line. The other lines of constant 
$\delta$, for fixed $k$, are also approximately parallel to the BFN line, and 
decreasing $\mid\delta\mid$ 
corresponds to entering into the PT region and going deep into it. All these 
lines concentrate onto the BFN line as $k$ approaches zero: for $k = 0$, 
$\delta =0$ 
except on the BFN line $1 + c_1 c_2 = 0$ where $\delta$ is undefined. We now
rephrase the conclusions above in terms of the three basic asymptotic states of
the CGLE in the PT regime. They will be valid as long as the phase 
description (\ref{kawae}) remains accurate. 
\begin{enumerate}
\item There are PT solutions of the quasiperiodic type for all values of the
parameters (as long as the phase description remains valid). Bounds on their 
velocity can be in principle obtained, but this is nontrivial since $Q$ 
is only known in an implicit way. 
\item Increasing $\mid\delta\mid$ by approaching the Eckhaus instability for a
given $k$ ($D_2=0$), or by 
increasing the
winding number reduces the variety of quasiperiodic solutions. 
\item Frozen turbulence solutions exist only for 
$\mid\delta\mid \lesssim 0.84$, that
is far enough from the line $D_2=0$ or for small enough winding number.  
\item There are linearly stable solutions in the main quasiperiodic branch for 
all values of the parameters. 
\item There are also riding turbulence attractors for all values of parameters.
\item For  $\mid\delta\mid > 3$, that is at high winding number or close 
enough to
the line $D_2=0$ the quasiperiodic solutions have a basin of attraction 
larger than
the riding turbulence ones.  
\end{enumerate}

A general feature of these conclusions is that the important quantity is
$D_2$, that is the distance in parameter space from the line at which the
$k$-wave became Eckhaus unstable. This line is {\sl below} the BFN line for $k
\neq 0$. Thus not only traveling waves, but also quasiperiodic, frozen 
turbulence,
and riding turbulence attractors should exist below
the BFN line for $k \neq 0$. In practice it is relatively easy to find
quasiperiodic 
states below but close the BFN line, but we have been unable  to find the other
two states so far. The difficulty in finding riding turbulence states can be a
consequence of the small range of winding numbers for which they are stable
($|\nu|=L |k|/(2\pi) < \nu_c$) so that the observability condition $\mid \delta
\mid < 3$ 
immediately brings us above the BFN line. Another possibility is that the
instability of the $\nu=0$ plane-wave attractor at the BFN
line has consequences of a global character beyond the validity of the phase
description.

The above predictions imply that the more promising zone for obtaining 
quasiperiodic solutions starting from random perturbations on a traveling wave 
of given winding number is for parameter values close and above $D_2=0$, or for 
high winding number 
($\mid\delta\mid >3 $). In any case no frozen turbulence 
should be observed in that zone. 

Some qualitative aspects of the conclusions above have been shown to be 
correct. In particular Torcini and collaborators \cite{torcini196,torcini296} 
have shown that the
average maximal Lyapunov exponent, quantifying the proportion of initial
conditions that fall into the spatio-temporal chaotic strange attractors, is a
decreasing function of $\nu$. 

Our numerical solutions also agree with the prediction  that quasiperiodic
 solutions show up more easily 
for small $D_2$. However, their basin of attraction appears to be much smaller 
than the implied by the conclusions of the phase description since it is 
reached with
very low probability from our initial conditions. This is specially true above
the BFN line. The reason for this is probably the
effect of the neglected term $D_{12}$, which is known to reduce the range of 
stable periodic solutions \cite{nepo195} and even to eliminate it by making 
the bifurcation subcritical \cite{janiaud1}. Above the BFN line the attractor 
that we observe more frequently at high winding number from our initial 
conditions is the frozen turbulent state. 

A more detailed checking of the predictions above would be desirable. This is 
however beyond the scope of the present paper since a detailed
theoretical analysis of the global properties of the phase space for the 
equation containing the term $D_{12}$ would be
probably needed beforehand. A promising alternative can be the study of the 
exact 
equation for $f(x-vt)$ in (\ref{uniftrans}) obtained in \cite{torcini296}. 

\section{Final Remarks}
\label{conclu}

One-dimensional wound-up phase turbulence has been shown to be much richer 
than the case $\nu=0$. 
The main results reported here, that is the existence of winding number
instability for phase-turbulent waves, the identification of the transition
PT-DT with the vanishing of the range of stable winding numbers, and the
coexistence of different kinds of PT attractors should in principle be observed
in systems for which PT and DT regimes above a Hopf bifurcation are known to
exist \cite{provansal94}. To our knowledge, there are so far no observations 
of the ordered PT states described above. There are however experimental
observations of what seems to be an Eckhaus-like instability for non-regular 
waves in the printer instability system \cite{debruyn94}. This suggests 
that the concept of a turbulent Eckhaus instability can be of interest beyond
the range of situations described by the CGLE. A point about which our study 
is inconclusive is the question on the existence of PT in the thermodynamic
limit. The identification of $\nu$ as an order parameter identifies the
continuation of Fig. \ref{fig7} towards larger system sizes as a way of
resolving the question. It should be noted however that although a linear
scaling of the order parameter with system size is usual in common phase 
transitions, broken ergodicity phase transitions, as the present one, generate 
usually a number of ordered phases growing exponentially with $L$, not
just linearly \cite{palmer89,mezard}. We notice   that the
results of Section \ref{asymstat} show that the states of a given $\nu$ are not
{\sl pure phases}, but different attractors are possible for given $\nu$.   
An order parameter more refined than $\nu$ should be able to distinguish 
between
the different attractors and, since some of them are disordered, the result of 
an exponentially large number of phases at large $L$ would be probably 
recovered. The results presented in Section \ref{kawa} give a justification for
the existence of the several wound states observed, an especific predictions
have been formulated on the basis of previous analitical and numerical results.  
Further work is needed
however to clarify the importance of the different terms in (\ref{exkseq}) 
and the validity of a phase description.

\section{ Acknowledgments}

Helpful discussions with L. Kramer, W. van Saarloos, H. Chat\'e, A. Torcini,
P. Colet and D. Walgraef are acknowledged. 
Financial support from DGYCIT (Spain) Projects PB94-1167 and PB94-1172 is
acknowledged. R.M. also acknowledges partial
support from the Programa de Desarrollo de las Ciencias B\'asicas (PEDECIBA,
Uruguay), and the Consejo Nacional de Investigaciones Cient\'\i ficas Y
T\'ecnicas (CONICYT, Uruguay). 

\appendix
\section{Numerical Integration Scheme}
\label{apend}
\noindent The time evolution of the complex field $A(x,t)$ 
subjected to periodic boundary conditions is obtained numerically
from the integration of the CGLE in Fourier space. The method
is pseudospectral and second-order accurate in time.
Each Fourier mode $A_q$ evolves according to:
\begin{equation}
\partial_t A_q(t) = - \alpha_qA_q(t) + \Phi_q(t) \>,
\label{Aqpunto}
\end{equation}
where $\alpha_q$ is $(1 + i c_1) q^2 - 1$, and
$\Phi_q$ is the amplitude of mode $q$ of the non-linear term in the CGLE.
At any time, the amplitudes $\Phi_q$ are calculated by
taking the inverse Fourier transform $A(x,t)$ of $A_q$, computing 
the non-linear term in real space and then calculating the direct 
Fourier transform of this term. A standard FFT subroutine 
is used for this purpose \cite{press89}.

Eq. (\ref{Aqpunto}) is integrated numerically in time by using a method similar
to the so called two-step method \cite{potter73}. For convenience in the
notation, the time step is defined here so as the time is 
increased by $2 \delta t$ at each iteration. 

When a large number of modes $q$ is used, the linear time scales
$\alpha_q$ can take a wide range of values. A way of circumventing this
stiffness problem is to treat exactly the linear terms by using the formal 
solution: 
\begin{equation}
A_q(t) = e^{-\alpha_qt} 
\left(
 A_q(t_0) e^{\alpha_q t_0} + \int_{t_0}^t \Phi_q(s) e^{\alpha_qs} ds
\right) \>.
\label{Sol}
\end{equation}
From here the following relationship is found: 
\begin{equation}
\frac{A_q(t + \delta t)}{e^{-\alpha_q\delta t}} -
\frac{A_q(t - \delta t)}{e^{ \alpha_q\delta t}} =
e^{-\alpha_qt} \int_{t-\delta t}^{t+\delta t} \Phi_q(s) e^{\alpha_qs} ds
\>.
\label{Aqnmas}
\end{equation}
The Taylor expansion of 
of $\Phi_q(s)$ around $s=t$ for small $\delta t$ gives an expression for 
the r.h.s. of Eq. (\ref{Aqnmas}):  
\begin{equation}
\Phi_q(t )
\frac{e^{\alpha_q\delta t} - e^{-\alpha_q\delta t}}{\alpha_q} +
{\cal O}(\delta t^3) \>.
\label{AkI}
\end{equation}
Substituting this result in (\ref{Aqnmas}) we get:
\begin{equation}
A_q(n+1) = e^{-2 \alpha_q\delta t} A_q(n-1) +
\frac{1 - e^{-2 \alpha_q\delta t}}{\alpha_q} \Phi_q(n) +
{\cal O}(\delta t^3) \>.
\label{tp3}
\end{equation}
where expressions of the form $f(n)$ are shortcuts for $f(t=n\delta t)$. 
Expression (\ref{tp3}) is the so called ''slaved leap frog" of Frisch et 
al.\cite{frisch86}. 
To use this scheme the values of the field at the first two time steps
are required. Nevertheless, this scheme alone is unstable for the CGLE. 
This is not explicitly stated in the literature and probably a corrective 
algorithm is also applied. 
Obtaining such correction is straightforward: Following steps
similar to the ones before one derives the auxiliary expression 
\begin{equation}
A_q(n) = e^{-\alpha_q\delta t} A_q(n-1) +
\frac{1 - e^{-\alpha_q\delta t}}{\alpha_q} \Phi_q(n-1) 
+ {\cal O}(\delta t^2) \>,
\label{tp2}
\end{equation}

The numerical method we use, which we will refer to as the
two--step method, provides the time evolution of the field 
from a given initial condition by using Eqs. (\ref{tp3}) and (\ref{tp2}) 
as follows: 
\begin{enumerate}
\item $\Phi_q(n-1)$ is calculated from $A_q(n-1)$ by going to real space. 
\item Eq. (\ref{tp2}) is used to obtain an approximation to $A_q(n)$. 
\item The non-linear term $\Phi_q(n)$ is now calculated from this $A_q(n)$  
by going to real space. 
\item The field at step $n+1$ is calculated from (\ref{tp3}) by using 
$A_q(n-1)$ and $\Phi_q(n)$. 
\end{enumerate}
At each iteration,
we get $A_q(n+1)$ from $A_q(n-1)$, and the time advances by $2 \delta t$.
Note that the total error is ${\cal O}(\delta t^3)$,
despite the error in the intermediate value obtained with Eq. (\ref{tp2}) is 
${\cal O}(\delta t^2)$. The method can be easily made exact for plane
waves (\ref{planew}) of wavenumber $k$ (and then more precise for 
solutions close to this traveling wave) 
simply by replacing the nonlinear term $\Phi_q$ in (\ref{Aqpunto}) by 
$\Phi_q + (1+ic_2)(1-k^2)A_{q}$, and subtracting the corresponding term
from $\alpha_q$. We have not implemented this improvement because we were 
mostly interested in solutions changing its winding number, so that they are
not close to the same traveling wave all the time.

The number of Fourier modes depends on the space discretization. 
We have used $dx=1$ and usually 
$N=512$. The time step was usually $dt=2 \delta t = 0.01$. The accuracy of the 
numerical method has been estimated by integrating plane-wave solutions. 
The amplitude and frequency of the field
obtained numerically will differ slightly from the exact amplitude and
frequency, not only due to round-off errors, but also due to the fact that
the method is approximate.
The method has been tested by using a stationary unstable traveling wave of wave
number $k$ as initial
condition. The numerical 
errors will eventually move the solution away from the plane--wave unstable 
state. To be precise, in a typical run with $c_1=-1.0$ and $c_2=2.4$, 
with $dt=0.01$ and $k=0.123$, the amplitude was kept constant 
to the fifth decimal digit during $\sim 8000$ iterations. In comparison, when 
a Gaussian noise with an amplitude as small as $10^{-7}$ is added to
the traveling wave, the modulus is maintained equal to
its steady value (up to the fifth decimal) during $1500$ iterations.
The frequency $\omega_q$ determined numerically by using $dt=0.01$ fits the
exact value up to the fourth decimal digit.

The integration method introduced here has also been applied succesfully
to the case of two coupled equations, or equivalently a Vectorial CGLE 
\cite{toni96,montagne96d}.


\begin{thebibliography}{10}
\bibitem{blowup}
A related question that also needs further study is the appearance of
  singularities at finite time in the solutions of (\protect\ref{exkseq})
  \protect\cite{chate8,janiaud1}.
\bibitem{shilnikov}
Simply stated Shil'nikov theory establishes that in a three dimensional
  dynamical system with a fixed point being a saddle-focus (i.e. unstable in
  one direction and oscillatory stable in the other) and provided there exists,
  for a value of the parameters, a trajectory homoclinic to the fixed point,
  the eigenvalues obtained through a linearization around the fixed point
  determine the global behaviour for this value of the parameters and a
  neighbourhood of it. If these eigenvalues are $\lambda_{1,2} = - \rho \pm i
 \omega \mbox{ and } \lambda_3 = \lambda$ ($\rho$,$\lambda>0$), then if $\rho >
   \lambda$ there is a countable infinity of chaotic trajectories near the
  homoclinic one. Proximity to the homoclinic trajectory implies that the shape
  of the chaotic trajectory is, as commented in the text, a sequence of
  irregularly spaced pulses. On the contrary, if $\rho < \lambda$ there are no
  chaotic trajectories. For a more precise statement of the theorems see
  \protect\cite{wiggins1,nayfeh1,kuznetsov}.
\end{thebibliography}

\end{document}